\documentstyle[12pt]{article}

\advance\hoffset -0.5cm
\advance\voffset -1.0cm
\newlength{\height}
\divide \height by 3

\begin{document}
%\pagestyle{empty}
%%%%%%%%%%%%%%%%%%%%%%%%%%%%%%%%%%%%%%%%%%%%%%%%%%%%%
\newcommand{\be}{\begin{equation}}
\newcommand{\bea}{\begin{eqnarray}}
\newcommand{\ee}{\end{equation}}
\newcommand{\eea}{\end{eqnarray}}
\newcommand{\Gze}{{\Gamma}_{\scriptstyle 0}}
\newcommand{\Gon}{{\Gamma}_{\scriptstyle 1}}
\newcommand{\Gtw}{{\Gamma}_{\scriptstyle 2}}
\newcommand{\gth}{{\gamma}_{\scriptstyle 3}}
\newcommand{\Gth}{{\Gamma}_{\scriptstyle 3}}
\newcommand{\Gfi}{{\Gamma}_{\scriptstyle 5}}
\newcommand{\gze}{{\gamma}_{\scriptstyle 0}}
\newcommand{\gon}{{\gamma}_{\scriptstyle 1}}
\newcommand{\gtw}{{\gamma}_{\scriptstyle 2}}
\newcommand{\gfi}{{\gamma}_{\scriptstyle 5}}
\newcommand{\Psib}{\overline \Psi}
\newcommand{\psib}{\overline \psi}
\newcommand{\emnl}{\epsilon_{\mu\nu\lambda}}
\newcommand{\uv}{U(1)_{\scriptstyle V}}
\newcommand{\ua}{U(1)_{\scriptstyle A}}
\newcommand{\alv}{\alpha_{\scriptstyle V}}
\newcommand{\ala}{\alpha_{\scriptstyle A}}
\newcommand{\all}{\alpha_{\scriptstyle L}}
\newcommand{\alr}{\alpha_{\scriptstyle R}}
\newcommand{\sth}{\Sigma_3}
\newcommand{\intdk}{\int {{\rm d}^3k\over (2\pi)^3}}
\newcommand{\dmn}{D_{\mu\nu}}
\newcommand{\magp}{\left|p\right|}
\def\spur#1{\mathord{\not\mathrel{#1}}} \baselineskip=\height

% for double spacing
\begin{titlepage}
\begin{center}
\makebox[\textwidth][r]{hep-th/9503131} \\
 \makebox[\textwidth][r]{SNUTP 95-022} \\
\makebox[\textwidth][r]{UFIFT 95-03}
\vskip 0.35in
{{\Large \bf Discrete Anomaly and Dynamical Mass in\\
2+1 dimensional $\uv\times\ua$ Model}}
\end{center}
\begin{center}
\par \vskip .1in \noindent Deog Ki Hong\footnote[1]
{E-mail address: dkhong@hyowon.cc.pusan.ac.kr}
\end{center}
\begin{center}
Institute for Fundamental Theory, University of Florida\\
Gainesville, FL 32611, U.S.A.\\
and\\
Department of Physics, Pusan National
University\footnote[2]{Permanant address}\\
Pusan 609-735, Korea\\
\par \vskip .1in
\noindent
\end{center}

\begin{abstract}
We note that in (2+1)-dimensional gauge theories with even number
of massless fermions, there is anomalous
$Z_2$ symmetry if theory is regularized in a parity-invariant way.
We then consider a parity invariant
$\uv\times\ua$ model, which induces a mutual Chern-Simons term in
the effective action due to $Z_2$ anomaly.
The effect of the discrete anomaly is studied in the induced spin
and in the dynamical  fermion mass.
%By solving the Dyson-Schwinger equation for fermion self energy,
%we find that in the large flavor ($N$) limit
%the fermion gets parity invariant mass dynamically when
%$N<N_c\simeq64/\pi^2$.

\vskip 0.2in
\noindent


\end{abstract}
\end{titlepage}
%\section{Introduction}
The symmetry of classical lagrangians often breaks
down upon quantization. A well-known example is
the axial anomaly in quantum electrodynamics \cite{jackiw},
where any gauge invariant regularization necessarily
breaks the axial symmetry.
On the other hand the irreducible spinor    representation of
Lorentz group in odd dimensions does not have $\gfi$-like object.
Namely, there is no matrix anti-commuting
with all $\gamma$ matrices in odd dimensions.
For instance, in three dimensions
the irreducible spinor     representation is two dimensional
and the product of all $\gamma$-matrices $\Gze\Gon\Gtw=1$.
Therefore there is no axial anomaly in odd diemsions.
But, a discrete symmetry might be anomalous in odd
dimensional gauge theories
due to the incompatibility of the gauge-invariant regulator with
the discrete symmetry.
The anomalous discrete symmetry is realized as an induced quantum
number for the vacuum \cite{wilczekgoldstone}.

Redlich \cite{redlich} has shown that parity is anomalous
in (2+1)-dimensional $SU(N)$ gauge theories since
the parity invariant regularization results in an effective action,
which is not invariant under large gauge
transformations, because $\Pi_3\left(SU(N)\right)=Z$
for $N\ge2$, and thus one needs a parity-violating Chern-Simons term
to recover the gauge invariance in the effective action.
For the abelian case,
parity is anomalous in perturbation theory \cite{redlich} and
for time-independent gauge fields
the parity anomaly can be understood as
the (1+1)D axial anomaly \cite{niemi}.

However, when the number of fermions
is even,
one can find a parity-preserving
Pauli-Villars regulator of four-component fermions
\cite{jackiw2, my2}.
Then,
parity is no longer anomalous and the Chern-Simons
term is not induced in the effective Lagrangian.
In this paper, we note that for even number of two-component fermions
there is another anomalous
discrete symmetry, which is not parity, and we study
the effect of this discrete anomaly in $U(1)_V\times U(1)_A$ model.
This model itself is also interesting since it might be realized in
parity-invariant planar superconductivity
\cite{mavromatos}.

The model is described by
\be
{\cal L} = -{1\over4} F_{\mu\nu}F^{\mu\nu}
-{1\over4} B_{\mu\nu}B^{\mu\nu}+
         \sum_{i=1}^{\scriptstyle N}{\overline \Psi}_i
         \gamma^{\mu} D_{\mu} \Psi_i,
\label{plagrangian}
\ee
where $\Psi_i$ is a four-component spinor made of a pair of
two-component spinors as

\be
\Psi_i(x)= \left(\begin{array}{c}
                      \psi_i(x) \\
                      \sigma^{3} \psi_{\scriptstyle N+i}(x)
                 \end{array}  \right)
\ee
(We consider even number of two-component massless spinors.)
The covariant derivative
$D_{\mu}=\partial_{\mu}-ieA_{\mu}-ig\gfi B_{\mu}$ and the field
strength tensors
$F_{\mu\nu}=\partial_{\mu}A_{\nu}-\partial_{\nu}A_{\mu}$,
$B_{\mu\nu}=\partial_{\mu}B_{\nu}-\partial_{\nu}B_{\mu}$.
The gamma matrices for the four-component spinors are defined as
\be
\gamma^0= \left(\begin{array}{cc}
                      0 & I            \\
                      I & 0
                \end{array}  \right ) ,
\quad
\gamma^i= \left(\begin{array}{cc}
                      0 & -\sigma^i            \\
                      \sigma^i & 0
                \end{array}  \right ) ,
\quad
\gamma^5= \left(\begin{array}{cc}
                      I & 0            \\
                      0 & -I
                \end{array}  \right ) .
\label{gammas}
\ee

Lagrangian (\ref{plagrangian}) has a global
$SU(N)\times SU(N)$ symmetry whose N\"other currents are
\be
J_{a}^{\mu}=\Psib\gamma^{\mu}T_a\Psi, \quad
J_{a5}^{\mu}=\Psib\gamma^{\mu}T_a\gamma^5\Psi
\label{ncurrents}
\ee
where $T_a$'s are the generators for $SU(N)$.
Note also that we impose parity and a discrete $Z_2$ symmetry to
forbid mass terms for fermions and gauge fields in the
Lagrangian (\ref{plagrangian}).
By forbiding the mass terms for fermion, we have
$U(1)_3\times U(1)_{35}$ ``chiral symmetry", generated by
$i\gamma^3$ and $\gamma^3\gamma^5$. This
``chiral symmetry", which mimics (3+1)-dimensional chiral symmetry,
is not really chiral symmetry but a part of the flavor symmetry
$U(2)$ for the two
two-component spinors constituting four-component spinors.
Parity $P$ is a space-time
transformation, $(t,x,y)\mapsto (t,-x,y)$, under which
fermion and gauge fields transform as
\bea
\Psi(t,x,y) & \stackrel{P}\longmapsto & \Psi^{\prime}(t,-x,y)
%              =i\gamma_0 e^{{i\over2}\pi\sigma_{23}} \Psi(t,x,y)
              =-\gamma^1\gamma^5\Psi(t,x,y) \\
(A_0(x),A_1(x),A_2(x)) & \stackrel{P}\longmapsto
       & (A_0(x), -A_1(x),A_2(x)) \\
(B_0(x),B_1(x),B_2(x)) & \stackrel{P}\longmapsto
       & (-B_0(x), B_1(x),-B_2(x)).
\eea
We see that $A_{\mu}$ and $B_{\mu}$ transform like an ordinary vector
and an axial vector, respectively.
Under $Z_2$ transformation,
\be
A_{\mu}\stackrel{Z_2}\longmapsto A_{\mu}, \quad
B_{\mu}\stackrel{Z_2}\longmapsto -B_{\mu},
\quad \Psi\stackrel{Z_2}\longmapsto i\gamma^3\Psi.
\label{discrete}
\ee
$Z_2$ tantamounts to the charge conjugation for $U(1)_A$,
the ``axial" coupling.

In the perturbation of (2+1)-dimensional gauge theories,
only the vacuum polarization and the triangle
graph are ultraviolet divergent. One may regularize the divergences
with the Pauli-Villars regulator. One has then two choices for the
regulator mass for each flavor $i$.
One is parity-invariant but $Z_2$-violating
($M\Psib_i\gamma^3\Psi_i$) and the other is $Z_2$-invariant but
parity-violating ($M\Psib_i\gamma^3\gfi\Psi_i$). Therefore
either parity or $Z_2$ (but not both) is anomalous, namely
$PZ_2$ is always anomalous.

Integrating out the
fermions, one gets $-i {\rm Tr} \ln i\spur{D}$ in the effective
action. In the perturbation theory, if one uses the parity-invariant
Pauli-Villars regulator,
one gets an effective Lagrangian given by
\be
{\cal L}_{\rm eff}=eg{N\over 2\pi}{M\over \left|M\right|}\emnl
B^{\mu}F^{\nu\lambda} +\cdots
\label{effective}
\ee
where $\cdots$ denotes the higher order terms and $M$ is the
regulator mass signifying the $Z_2$ anomaly.
The leading term in the effective Lagrangian (\ref{effective})
can be obtained from the Feynman diagram in Fig. 1.
This term is similar to the Chern-Simons term but it couples two
different gauge fields. We call this a mutual Chern-Simons term.
It leads to mutual fractional statistics and is believed to be
realized in a layered Hall system
exhibiting a filling factor of even denominator \cite{wilczek}.
One can see easily that the mutual Chern-Simons term in
$\uv\times\ua$ theory is the only term in perturbation theory which
breaks $Z_2$ in the effective action.
Had we chosen $Z_2$-invariant Pauli-Villars regulator, we would have
gotten Chern-Simons terms for each gauge  fields breaking
parity.

The radiative generation of the mutual Chern-Simons term is also
noted in references \cite{mavromatos},
where they argued
that $\uv\times\ua$ arises in a model of dynamical holes in a
planar quantum antiferromagnet in the large spin and small doping
limit.
But, here, we point out the origin of the mutual Chern-Simons
term in (\ref{effective}) is $Z_2$ anomaly and we argue that
one can not avoid it in parity-invariant
theories in 2+1 dimensions because
the parity-invariant regulator necessary breaks $Z_2$.

Due to the mutual Chern-Simons term,
fermions get a fractional spin  $s={1\over N}$ by the usual
Aharanov-Bohm effect \cite{wilczek1}.
At long distances a particle
carrying unit (axial) charge $g$ will look like a localized
vortex of magnetic flux $\Phi=2\pi/eN$ (modulo a sign which is
not important here) for a particle of unit (vector) charge $e$.
Therefore a fermion orbiting around another fermion will get
a   Aharanov-Bohm phase $e\Phi$ and thus the induced spin
$s=e\Phi/2\pi={1\over N}$.
The spin of the
four-component spinors is invariant under parity:
\be
s=\int {\rm d}^2x \Psi^{\dagger}
  {i\over4}\left[\gamma_1,\gamma_2\right]\Psi
  \stackrel{P}\longmapsto
\int {\rm d}^2x \left(\gamma_1\gamma_5 \Psi\right)^{\dagger}
  {i\over4}\left[\gamma_1,\gamma_2\right]
  \left(\gamma_1\gamma_5\Psi\right)=s
\ee
The induced spin for a four-component fermion therefore does not
break parity. This is not the case for the two-component
fermion which can have only one direction for spin,
while the four-component fermion has two two-component spinors
which have spins of opposite direction.
The parity-violating Chern-Simons term
affects the dynamical generation of parity-even
mass for fermion in a rather interesting way \cite{hong, kondo}.
It tends to break parity maximally. Namely,
it reduces both of critical flavor number for mass generation
and the magnitude of mass itself. We study how       the
(radiatively generated) mutual Chern-Simons term affects
the dynamical generation of parity-even fermion mass.
According to a general theorem by Vafa and Witten
\cite{vafa, park}, parity-odd
fermion mass cannot      be generated dynamically in a
parity-invariant $ \uv \times \ua$ model.
We use the $1/N$ expansion,
since it not only gives a systemmatic way of treating
nonperturbative phenomena but also softens the IR divergences of
perturbative three-dimensional gauge theories \cite{pisa}.
To have a well-defined field theory in large flavor ($N$)
limit, we keep $\alv\equiv e^2N$ and $\ala\equiv g^2N$ finite as $N$
goes to infinity.

In     leading order in $1/N$ expansion, the gauge-boson propagators
get contribution from the fermion loops and they get mixed.
They are
\bea
D_{\mu\nu}^{AA}(p)\!\!\!\! & = &\!\!\!\!
        {  -i(g_{\mu\nu}-p_{\mu}p_{\nu}/p^2) \over p^2  }
        {   1+ {1\over\pi}\ala F   \over
            (1 + {1\over\pi}\ala F)(1+ {1\over\pi}\alv F)
        + {m^2\over4\pi^2} \ala\alv (G/p^2)^2       }, \\
D_{\mu\nu}^{AB}(p)\!\!\!\! &  =  & \!\!\!\!
        {  \epsilon_{\mu\nu\lambda} p^{\lambda} \over p^2   }
        { { m\over2\pi}\kappa  G/p^2     \over
           (1+ { 1\over\pi} \ala F)(1+ { 1\over\pi}\alv F)
           + {m^2\over4\pi^2} {\kappa}^2 (G/p^2)^2        },   \\
D_{\mu\nu}^{BA}(p)\!\!\!\!  &   =  &
\!\!\!\!  D_{\mu\nu}^{AB}(p),  \\
D_{\mu\nu}^{BB}(p)\!\!\!\!  &  =   &
\!\!\!\! D_{\mu\nu}^{\rm AA}(p)
(\ala \leftrightarrow \alv),
\eea
where the superscript ${\scriptstyle AB}$
means gauge fields $A_{\mu}$ propagate to
gauge fields $B_{\mu}$ etc. and $\kappa=\sqrt{\ala\alv}$.
The functions $F$ and $G$ are
\bea
F(m^2, p^2)=\int_0^1 dx { x(1-x)\over \sqrt{m^2-x(1-x) p^2}  } \\
G(m^2, p^2)=\int_0^1 dx { 1\over \sqrt{m^2-x(1-x) p^2}  },
\eea
which come from the one-loop vacuum polarization.
To calculate the vacuum polarization,
we need to know the exact form of the fermion self-energy, which
requires full solutions to the Dyson-Schwinger equations. As
an approximation, we take a constant mass for the self-energy,
$\Sigma(p)=m\gamma^3$, which
must be very small compared to the scale of the theory, $\ala$ or
$\alv$, since it is generated by a nonperturbative $1/N$ effect, and
it must be also parity-even. Note also that
by the $U(1)_3\times U(1)_{35}$ ``chiral symmetry" one can always
rotate the fermion self energy to be
$\Sigma(p)=\gamma_3\Sigma_3(p)$, where $\Sigma_3(p)$ is a function
proportional to the unit matrix.

In $1/N$ perturbation, the full vertex function can be expanded as
\be
\Gamma_{\mu}=\gamma_{\mu}+O({1\over N}).
\ee
For the leading order, we take $\Gamma_{\mu}=\gamma_{\mu}$.
Then, the Ward-Takahashi identity requires
the wave-function renormalization constant to be 1 for a consistent
$1/N$ expansion.
The Dyson-Schwinger gap
equation (Fig. 2) in Euclidean notation is then
\bea
\gamma_3\Sigma_3(p)\!\!\!&=&\!\!\!{\alv\over N}
\intdk D^{AA}_{\mu\nu}(p-k)\gamma_{\nu}
 { \spur{k}-\gth\sth(k) \over  k^2 + \sth^2(k)  }  \gamma_{\mu}
        \nonumber \\
\!\!\! & &\!\!\!\!\!\!\!\!+{\kappa\over N} \intdk \left (
        \dmn^{AB}\gamma_{\mu}
 { \spur{k}-\gth\sth(k) \over  k^2 + \sth^2(k)  }  \gfi\gamma_{\mu}
       + \dmn^{BA}\gfi\gamma_{\mu}
    { \spur{k}-\gth\sth(k) \over  k^2 + \sth^2(k)  }  \gamma_{\mu}
    \right) \label{Dyson} \\
\!\!\! & &\!\!\!\!\!\!\!\!  + {\ala\over N}
        \intdk D^{BB}_{\mu\nu}(p-k)\gfi\gamma_{\nu}
 { \spur{k}-\gth\sth(k) \over  k^2 + \sth^2(k)  } \gfi\gamma_{\mu}.
 \nonumber
\eea
Since the dynamically generated mass $m$ is exponentially small
compared to the scale, $\ala$ and $\alv$, and the (2+1)-dimensional
gauge theories are superrenormalizable, one can think of $m$ as an
infrared cutoff and $\alv$ (or $\ala$) as a ultraviolet cutoff.
For the momentum $p$ in $m<p<\alv$ or $\ala$, one can simplify the
expression for the vacuum polarization tensor. Namely, for momentum
for this range,
\be
        F(m^2, p^2)\simeq {  \pi\over8\magp  },
        \quad G(m^2, p^2)\simeq  { \pi\over\magp },
\ee
and the gauge fields propagators in Euclidean space are
\bea
\dmn^{AA}(p)\!\!\!&\simeq&\!\!\!
        {g_{\mu\nu}-p_{\mu}p_{\nu}/p^2 \over p^2 }
        { \magp \over 8\alv  },  \\
\dmn^{BB}(p)\!\!\!&\simeq&\!\!\!
        {g_{\mu\nu}-p_{\mu}p_{\nu}/p^2 \over p^2 }
        { \magp \over 8\ala  },  \\
\dmn^{AB}(p)\!\!\!&\simeq\!\!\!&32{ m \over \kappa }
        { \epsilon_{\mu\lambda\nu} p^{\lambda}\over \magp^3  }.
\eea
We see that the propagator $\dmn^{AB}$ is proportional to
$m/\kappa$ while the other propagators are the ones for
$m\rightarrow0$. Though by dimensional counting
$\dmn^{AB}$ is quite suppressed compared to other propagators, it is
not clear that one can neglect the second term in (\ref{Dyson}).
However, if one analyzes the Dyson-Schwinger equation,
keeping the second term,
one finds at the end that keeping the second term is equivalent to
adding a constant mass to $\sth(p)$. Therefore it is not
consistent with the massless limit approximation
($m\rightarrow0$) taken for $\dmn^{AA}$
and $\dmn^{BB}$, if one keeps the second term in (\ref{Dyson}) which
is proportional to $m$.

With the second term dropped, the Dyson-Schwinger equation
(\ref{Dyson}) becomes exactly same as that of pure $QED_3$
analyzed by many other people \cite{dynamass, appelquist},
except that now there are two copies of gauge fields.
The analysis goes parallel to the analysis in
\cite{appelquist}. Here we present the result in a slightly different
fashion, following the analysis by Cohen and Georgi for
(3+1)-dimensional gauge theories in the ladder
approximation \cite{georgi}, where the physical meaning of constants
appearing in the
asymptotic behavior of the fermion self energy is identified with
the operators in the operator product expansion of the fermion
two-point function.

Taking the trace over $\gamma$ matrices after
mutiplying $-\gth$ and performing the angular integral in
(\ref{Dyson}), we get
\bea
\sth(p)={\alv\over2\pi^2 Np}\int {\rm d}k{k\sth(k)\over
       k^2+\sth^2(k)}\ln{k+p+\alv/8\over\left|k-p\right|+\alv/8}
       \nonumber \\
   + {\ala\over2\pi^2 Np}\int {\rm d}k{k\sth(k)\over
       k^2+\sth^2(k)}\ln{k+p+\ala/8\over\left|k-p\right|+\ala/8}.
\label{angular}
\eea
As was done in \cite{appelquist}, we expand the logarithm in power
series for $p\ll \alpha$ (here $\alpha\simeq\alv$ or $\ala$ is
a typical scale of the theory) to get
\be
\sth(p)={8\over \pi^2 Np} \int {\rm d}k
        {k\sth(k) \over k^2+\sth^2(k) }
        \left( p+k-\left|p-k\right| \right).
\label{angle}
\ee
Differentiating the integral equation (\ref{angle}),
we obtain
\be
\sth^{\prime}(p)=-{16\over \pi^2 N} \int_0^p { {\rm d}k \over p^2 }
        { k^2\sth(k) \over k^2+\sth^2(k) },
\label{diff1}
\ee
where $^{\prime}$ denotes differentiation with respect to $p$.
We see from (\ref{diff1})
\be
\lim_{p\rightarrow0}p^2\sth^{\prime}(p)=0
\label{boundary1}
\ee
which serves as an infrared boundary condition for $\sth(p)$.
On the other hand, the equation we get by
differentiating after multiplying $p$
\be
\left(p\sth\right)^{\prime}=-{16\over \pi^2 N}\int_p^{\alpha}
{\rm d}k {k\sth(k)\over k^2+\sth^2(k)}
\ee
gives an ultraviolet boundary condition
\be
\lim_{p\rightarrow\alpha}\left(p\sth\right)^{\prime}=0
\label{ultra}
\ee
Mutiplying by $p^2$ and differentiating once again we obtain
\be
        p^2\sth^{\prime\prime}+2p\sth^{\prime}+
        {r\over4} {p^2\sth \over p^2+\sth^2}=0
\label{diff2}
\ee
where $r=N_c/N$ with $N_c=64/\pi^2$.
For small $p$, the solution to (\ref{diff2}), which is consistent
with the boundary condition (\ref{boundary1}) is
\be
\sth(p)=m_C\qquad{\rm for}\qquad p\ll\sth(p),
\label{small}
\ee
and for large $p$ ( $p\gg\sth(p)$ )
\be
\sth(p)=m_R\left({p\over\mu}\right)^{-\epsilon}+
        {\kappa\over p}\left({p\over\mu}\right)^{\epsilon}
\label{large}
\ee
where
\be
\epsilon={1-\sqrt{1-r}\over2}.
\ee
and $\mu$ is the renormalization point.
As was shown in \cite{georgi}, the parameters $m_R$ and $\kappa$
correspond to a renormalized mass and a fermion condensate
$\left<\Psib\gth\Psi\right>$, respectively.
If $N>N_c$, one finds that $m_C$ has to be zero
in the chiral limit ($m_R\rightarrow0$), and thus $\sth(p)=0$.
Dynamical mass is not generated
and the trivial vacuum is the only solution \cite{georgi}.
When $N<N_c$, the solution to (\ref{diff2}) is
\be
\sth(p)={ A\over\sqrt{p} }\cos\left(\sqrt{r-1}\ln(p/\mu)+\phi\right),
\label{sol1}
\ee
where $A$ and $\phi$ are arbitrary constants.
We see that the operators $m_R{\bf 1}$  and $\Psib\gth\Psi$ are
coalesced due to strong interaction when $N<N_c$
and can not be distinguished
by the operator product expansion.
{}From the Dyson-Schwinger equation (\ref{Dyson}), we know that
for high momentum $p>\alpha$
\be
\sth(p)\simeq {C\over p^2}
\label{sol2}
\ee
At $p\simeq\alpha$ the solution (\ref{sol1}) for $p<\alpha$
should be smoothly connected to
the solution (\ref{sol2})
for $p>\alpha$. This condition is given by
the boundary condition (\ref{ultra}) at $p=\alpha$.
Taking the renormalization point to be $\mu\simeq m_C$,  we get
\be
m_C=\alpha e^{ -{\pi-\phi \over \sqrt{N_c/N-1} } },
\label{mass}
\ee
where $N_c=64/\pi^2$.
We see that the dynamical mass generation in $\uv\times\ua$ is
precisely same as pure $QED_3$ except that the critical flavor
is now doubled.

Since $Z_2$ is anomalous, one may start with a
bare mutual Chern-Simons
term in this $\uv\times\ua$ model:
\be
{\cal L}^{\prime}={\cal L}+
        {\kappa_0\over2\pi} \emnl B^{\mu}F^{\nu\lambda}.
\ee
Then the $Z_2$-violating (but parity-even) fermion mass will be
generated radiatively in perturbation theory. However, we can still
ask whether this $Z_2$-violating mutual Chern-Simons term will
affect the dynamical generation of parity-even (namely $Z_2$
violating) fermion mass. (The parity-odd mass is not generated,
even in nonperturbative analysis,
whether the mutual Chern-Simons term is present or not.)
The analysis is again done by solving the Dyson-Schwinger equation
in $1/N$ expansion.
For $m\ll p\ll \alpha$ or $\kappa_0$, the Dyson-Schwinger
equation will look same as before except now the propagator
$D_{\mu\nu}^{AB}$ is no longer negligible:
\be
D_{\mu\nu}^{AB}(p)={\epsilon_{\mu\lambda\nu}p^{\lambda}\over p^2}
        { {\pi\over 2\kappa}\over \left(1+{\ala\over 8\magp}\right)
        \left(1+{\alv\over8\magp}\right) }.
\ee
With $D_{\mu\nu}^{AB}$ the Dyson-Schwinger equation becomes, after
performing the angular integration,
\be
\sth(p)={8\over \pi^2 Np} \int {\rm d}k
        {k\sth(k) \over k^2+\sth^2(k) }
        \left( p+k-\left|p-k\right| \right)
        -{1\over N\pi}{\kappa\over\kappa_0}\int {\rm d}k
        {k^2\over k^2+\sth^2(k)},
\label{bare1}
\ee
where we keep only the leading term in $p/\alpha$.
We see that the mutual Chern-Simons term contributes to $\sth(p)$ by
a constant, which is same as having a bare mass term in
the Lagrangian. Therefore, the leading contribution of the bare
mutual Chern-Simons term is radiative generation of $Z_2$ violating
(parity-even) fermion mass. It does not affect the nonperturbative
generation of fermion mass.

If one transforms the gauge fields into new ones as
\be
A_{\mu}=a_{\mu}+b_{\mu},\qquad
B_{\mu}=a_{\mu}-b_{\mu},
\ee
the mutual Chern-Simons term becomes
\be
{\kappa_0\over2\pi}\epsilon_{\mu\nu\lambda}B^{\mu}F^{\nu\lambda}=
{\kappa_0\over2\pi}\epsilon_{\mu\nu\lambda}a^{\mu}a^{\nu\lambda}-
{\kappa_0\over2\pi}\epsilon_{\mu\nu\lambda}b^{\mu}b^{\nu\lambda},
\ee
where
$a^{\nu\lambda}=\partial^{\nu}a^{\lambda}-\partial^{\lambda}a^{\nu}$
and
$b^{\nu\lambda}=\partial^{\nu}b^{\lambda}-\partial^{\lambda}b^{\nu}$.
And the covariant derivative becomes
\be
        D_{\mu}=\partial_{\mu}
        -{e+g\gfi\over2}a_{\mu}-{e-g\gfi\over2}b_{\mu}.
\ee
The gauge fields $a_{\mu}$ and $b_{\mu}$ decouple at tree level, but
they get coupled through fermion loops.
When $e=g$,
$a_{\mu}$ and $b_{\mu}$ become the gauge fields for
$U(1)_L$ and $U(1)_R$, generated by $(1+\gfi)/2$ and
$(1-\gfi)/2$, respectively.
The upper two-component spinor has $U(1)_L$ charge $e$
but no $U(1)_R$ charge and
the lower two-component spinor has $U(1)_R$ charge $e$
but no $U(1)_L$ charge. They are completely decoupled. In this case
$\uv\times\ua$ model is just two copies of $QED_3$ with a
Chern-Simons term of opposite sign and
$N$ two-component spinors. They are related by parity.
Under the parity, $a_{\mu}$ transforms to $b_{\mu}$, the upper
two-component spinor in a four-component spinor transforms to the
lower two-component spinor, and vice versa.
The symmetry is but still $U(N)\times U(N)\times P$.
One interesting is that, when fermion gets dynamical mass,
$U(N)\times U(N)$ breaks down to
$U(N/2)\times U(N/2)\times U(N/2)\times U(N/2)$
for even $N$, which is shown to occur in $1/N$ expansion
when $N<N_c/\left[1+(16\kappa_0/\alpha)^2\right]$
with $\alpha=e^2N$ \cite{hong}.

%\section {conclusion}
In conclusion,
we see that for an even number of two-component fermions in
(2+1)-dimensional gauge theories
$Z_2$ is anomalous if one
regularizes theory in a parity-invariant way.
Due to $Z_2$ anomaly a parity invariant $\uv\times\ua$
theory induces a mutual Chern-Simons term in the effective action,
which leads to fractional spin to fermions in the theory.
But, the radiatively generated mutual Chern-Simons term does not
affect the dynamical generation of fermion mass at least in the
leading order in $1/N$ expansion.
Fermions get dynamical mass when $N<64/\pi^2$ as if we have
two copies of three dimensional $QED$.
When a bare mutual Chern-Smions term is added, $Z_2$ violating
fermion mass is generated radiatively but the nonperturbative
generation of fermion mass does not get affected.

\vskip .1in
\noindent

{\bf Acknowledgments}
\vskip .1in

This work was supported in part by
the
Korea Science and Engineering Foundation through SRC program of
SNU-CTP, by NON DIRECTED RESEARCH FUND,
Korea Research Foundation, and also by Basic Science Research
Program, Ministry of Education, 1994 (BSRI-94-2413).
The author is
grateful to Prof. P. Ramond for reading the manuscript carefully
and he
would like to thank the high energy theory group at
   the Institute of Fundamental Theory,
University of Florida for its support and warm hospitality.
\pagebreak

\section*{Figure Captions}
\typeout{Figure Captions}
\begin{description}
\item[Figure 1:]
$Z_2$ anomaly.
The solid lines denote fermions,  the wavy lines gauge fields.
\item[Figure 2:] Dyson-Schwinger gap equation. The (bold) solid lines
denote (full) fermion propagator, the wavy lines gauge fields.
\end{description}


\begin{thebibliography}{99}

\bibitem{jackiw}S. Adler, Phys. Rev. {\bf 177}, 2426 (1969);
    J. S. Bell and R. Jackiw, N. Cimento {\bf 60A}, 47 (1969).

\bibitem{wilczekgoldstone} J. Goldstone and F. Wilczek,
    Phys. Rev. Lett. {\bf 47}, 986 (1981);R. Jackiw and C. Rebbi,
    Phys. Rev. {\bf D13}, 3398 (1976).

\bibitem{redlich}A. N. Redlich, Phys. Rev. Lett. {\bf 52}, 18 (1984);
     Phys. Rev. {\bf D29}, 2366 (1984).

\bibitem{niemi}A. J. Niemi and G. W. Semenoff, Phys. Rev. Lett.
        {\bf 51}, 2077 (1983).

\bibitem{jackiw2}R. Jackiw and S. Templeton,
   Phys. Rev. {\bf D 23}, 2291 (1981);
   S. Deser, R. Jackiw, and S. Templeton,
   Ann. Phys. {\bf (NY) 140}, 372 (1982);
   {\bf (E) 185}, 406 (1988);
   Phys. Rev. Lett. {\bf 48}, 975 (1982);
   {\bf (C) 59}, 1981 (1987);
   G.W. Semenoff and L.C.R. Wijewardhana,
   {\it ibid. } {\bf 63}, 2633 (1989).

\bibitem{my2}D. K. Hong and S. H. Park,
   Phys. Rev. {\bf D49}, 5507 (1994).

\bibitem{mavromatos}N. Dorey and N. Mavromatos, Nucl. Phys.
        {\bf B386}, 614, (1992); Phys. Lett. {\bf B250}, 107 (1990);
        G. W. Semenoff and N. Weiss,
        Phys. Lett. {\bf B250} 117 (1990);
        T. Banks and J. D. Lykken, Nucl. Phys. {\bf B336},
           500 (1990).

\bibitem{wilczek}F. Wilczek, Phys. Rev. Lett. {\bf 69}, 132 (1992).

\bibitem{wilczek1}F. Wilczek, in
        {\it Fractional Statistics and Anyon Superconductivity},
        edited by F. Wilzcek
        (World Scientific, Singapore 1990).

\bibitem{hong}D. K. Hong and S. H. Park, Phys. Rev.
   {\bf D 47}, 3651 (1993).

\bibitem{kondo}T. Ebihara, T. Iizuka, K.-I. Kondo
   and E. Tanaka, Chiba University Preprint CHIBA-EP-77
   (unpublished), hep-ph/9404361; K.-I. Kondo and P. Maris,
   Chiba/Nagoya Univ. Preprint, CHIBA-EP-84/DPNU-94-33,
   hep-ph/9408210; K.-I. Kondo and P. Maris, Chiba/Nagoya Univ.
   Preprint, CHIBA-EP-85/DPNU-94-51, hep-ph/9501280.

\bibitem{vafa}C. Vafa and E. Witten
  {\it Nucl. Phys.} {\bf B234}, 173 (1984).

\bibitem{park}S. H. Park and Y. Shamir, Phys. Rev.
        {\bf D48}, 3352 (1991).

\bibitem{pisa}T. Appelquist and R. D. Pisarski, Phys. Rev.
        {\bf 23}, 2305 (1981); R. Jackiw and S. Templeton,
        Phys. Rev. {\bf 23}, 2291 (1981).

\bibitem{dynamass}T. Appelquist, M. Bowick, E. Cohler, and L. C. R.
        Wijewardhana, Phys. Rev. Lett. {\bf 55}, 1715 (1985);
        T. Appelquist, M. Bowick, D. Karabali,
   and L. C. R. Wijewardhana, Phys. Rev. {\bf D 33}, 3704 (1986);
   R. Pisarski, {\it ibid.} {\bf D 29}, 2423 (1984);
   T. Appelquist, M. Bowick, D. Karabali, and L. Wijewardhana,
{\it ibid.} {\bf D 33}, 3774 (1986);D. Boyanovsky, R. Blankenbecler,
   and R. Yahalom, Nucl. Phys. {\bf B270}, 483 (1986);
   S. Rao and R. Yaholom, Phys. Rev. {\bf D34}, 1194 (1986);
   K. Stam, {\it ibid.} {\bf 34}, 2517 (1986).

\bibitem{appelquist}T. Appelquist, D. Nash and L.C.R. Wijewardhana,
        Phys. Rev. Lett. {\bf 60}, 2575 (1988); D. Nash, {\it ibid.}
        {\bf 62}, 3024 (1989); T. Appelquist and D. Nash, {\it ibid.}
        {\bf 64}, 721 (1990).


\bibitem{georgi}A. Cohen and H. Georgi, Nucl. Phys. {\bf B314}, 7
(1989).

\end{thebibliography}
\end{document}